# Influence of Columnar Defects on Magnetic Relaxation of Microwave Nonlinearity in Superconducting YBCO Resonator Devices

A.R. Medema[a], G. Ghigo[b,c], and S.K. Remillard[a,+]

[a] Hope College, Holland, MI, 49423, USA
[b] Politecnico di Torino, Dept. of Applied Science and Technology, c.so Duca degli Abruzzi 24,10129 Torino, Italy
[c] Istituto Nazionale di Fisica Nucleare, Sezione di Torino, via P. Giuria 1, 10125 Torino, Italy
[+] Corresponding author, remillard@hope.edu, +1-616-395-7507

## Abstract

Intermodulation distortion (IMD) of microwave signals incident upon a superconducting $YBa_2Cu_3O_{7-x}$ thin film sample on a $LaAlO_3$ substrate is shown to relax as the sample magnetization also relaxes, revealing the signature of fluxon decay in the microwave distortion. The sample contained a microchannel of columnar defects that was introduced using heavy ion beam irradiation. After exposure to and then removal of a low strength static magnetic field, IMD relaxed due to thermal activation of fluxons over the surface barrier. This process influenced second and third order IMD differently, visible in the differing functional form of the IMD in time and the temperature dependence of the relaxation function. The presence of columnar defects in the test sample suppressed thermal activation by pinning of fluxons, notably with complete suppression of relaxation in third order IMD at low temperatures. Higher order IMD was found to relax with the same functional form as the second and third order IMD of the same parity.

## 1. Introduction

High temperature superconducting microwave signal filters offer superior performance over conventional filters in both military and telecommunications applications but are affected by nonlinearity of the superconducting material [1]. These devices are also sensitive to external static magnetic fields, which degrade performance by increasing surface resistance. Intermodulation distortion (IMD) is influenced by the presence of fluxons, and in this paper it is shown that IMD responds to the fluxon density as demagnetization occurs.

In work by other authors, second order nonlinearity in cuprate high temperature superconductors (HTS) has been attributed to rectification of the flux lattice [2], and third order microwave nonlinearity has been attributed to the nonlinear Meissner effect [2,3]. Nonlinearity of the same parity (even or odd) is generally understood to be caused by the same physical source [4].

Trapped in potential wells at material defects, fluxons can become unpinned by thermal activation [5]. The time for depinning to occur varies with well depth $U$ and temperature $T$ as $t \sim e^{U/k_B T}$, where $k_B$ is the Boltzmann constant. In the Anderson-Kim model, a logarithmic time decay of the density of pinned fluxons emerges from the linear relationship between $U$ and current density [5]. Logarithmic relaxation also results from thermal activation over the surface barrier in the Bean-Livingston model [6]. The surface barrier in a finite superconducting sample results from competition between the attractive image force and the repulsive field gradient. Material defects are not required for this, only a potential barrier at the surface of the superconductor.

Irreversibility from depinning prompts hysteresis in superconductors [7], which corresponds to relaxation of the magnetization and microwave nonlinearity as flux is slowly expelled [8]. Following the removal of an external magnetic field, nonlinearity relaxes to a new value. A variety of time dependencies has been found to describe the relaxation and to selectively influence different orders of nonlinearity. Second order nonlinearity decays by an exponential fast process on the order of 10 seconds or less, followed by a slower logarithmic process. The fast process is attributed to repulsive diffusion of fluxons after removal of the field, and the logarithmic process is attributed to thermal activation over the surface barrier [8,9]. Third order nonlinearity is observed to relax predominantly by the slow logarithmic process, which masks the fast exponential decay that is more clearly seen in second order. Third order IMD exhibits a smaller overall change in signal strength than does second order IMD [9].

Engineered columnar defects have been previously examined for their local enhancement of intermodulation distortion [10], for changing local $T_c$ and critical current, for controlling second harmonic emission [11], and for their response to a magnetic field [7]. This parallels other recent work on iron-based superconductors using both heavy ions [12,13] and lower energy proton beams [14,15]. By providing a pinning center with similar geometry and thus enhancing the activation energy for flux creep, columnar defects reduce the strength of magnetic relaxation [7]. This paper replicates and expands upon the previous results in reference [9] with a wider temperature range and higher time resolution, investigates the effects of the added columnar defects for controlling magnetic relaxation in superconducting devices, and examines relaxation in higher order nonlinearity.

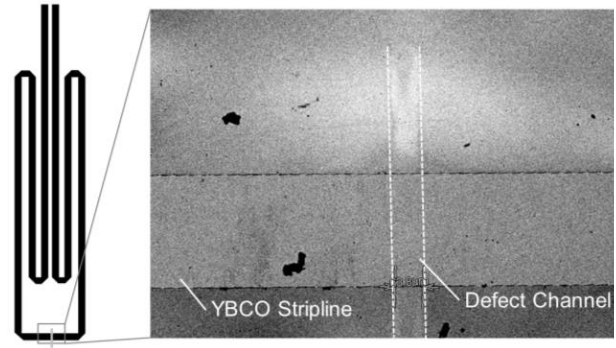

**Figure 2.** Scanning electron microscope image of the channel of columnar defects. The two dashed lines are 57 μm apart.

## 2. Experiment

This paper describes a comparative study of two identical hairpin microwave resonators made from approximately 400 nm thick superconducting YBa$_2$Cu$_3$O$_{7-x}$ (YBCO) on a LaAlO$_3$ substrate. Each device under test was a thin film microstrip, resonant at 830 MHz, made with dimensions and fabrication methods described in references [10] and shown in Figure 1.

A 57 μm-wide channel of columnar defects was implanted perpendicularly to the superconducting surface using a beam of 250 MeV Au ions in the test sample ($T_C$=88.2 K) shown in Figure 2. The identical control sample ($T_C$=89.8

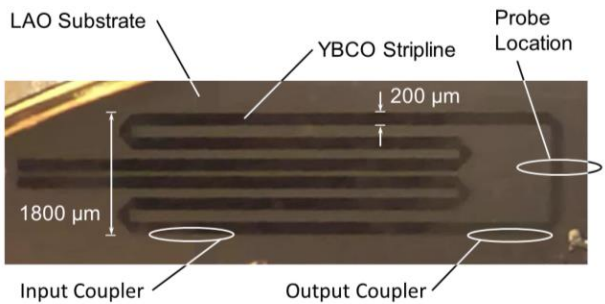

**Figure 1.** Dimensions and RF testing arrangement for each resonator sample

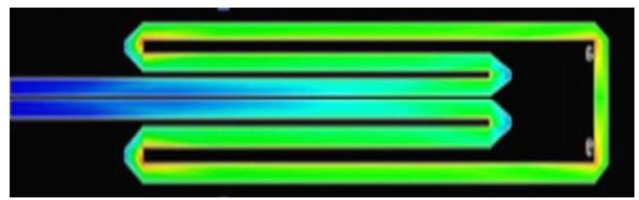

**Figure 3.** Microwave current distribution in the device under test simulated using method-of-moments IE3D

K) did not have this modification. The current distribution was simulated using IE3D [16] and is shown in Figure 3. The highest current is on the right side of the figure, midway between the two free ends of the structure, which are on the left side.

A block diagram of the microwave setup is shown in Figure 4. The signal mixing technique described in [1] was used to generate intermodulation distortion (IMD). A driving in-band tone $f_3$ was introduced with a near field antenna. A second antenna coupled signal out of the resonator sample to a Keysight CXA signal analyzer (the "spectrum analyzer") for measurement of transmitted signals in the resonator. Two out-of-band signals $f_1$ and $f_2$ were combined before being coupled into the resonator by a near field loop probe. IMD is generated locally by the sample nonlinearity at the point of the probe.

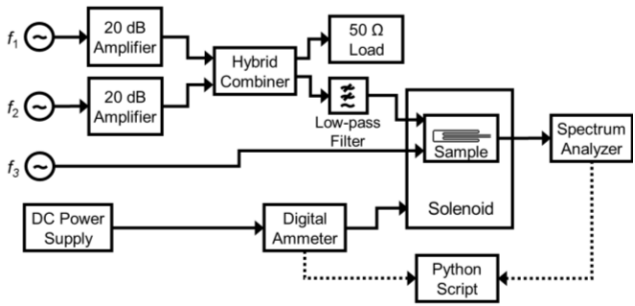

**Figure 4.** Block diagram of the automated intermodulation distortion decay measurement.

Figure 1 indicates the locations of the two coupling antennae. Also indicated is the location of the probe near the sample surface that excites in-band IMD using low frequency tones. Mixing of the input frequencies in the superconductor generates second order IMD at frequencies $f_{IM2} = f_3 \pm f_1$ and third order IMD at frequencies $f_{IM3} = f_3 \pm (f_2 - f_1)$ [1].

Sample resonance was excited by the signal at $f_3$, which varied from 790 MHz to 840 MHz depending on temperature. $f_1$ and $f_2$ were set to 50 kHz and 249 kHz, respectively, producing IMD inside the resonant band of the sample. The driving signal at $f_3$ was set to +15 dBm, while the probe signals $f_1$ and $f_2$ were each set to +23 dBm.

Time relaxation of IMD was measured at fixed temperature in a liquid nitrogen cooled cryostat contained inside a dewar, both fabricated from non-magnetic stainless steel. The cryostat was pressurized to 1 atm with helium exchange gas. Sample temperature was maintained by a resistive heater and temperature probe wired to a LakeShore 330 temperature controller.

Measurements were controlled by a Python script with user-defined values for the instrument precision and time resolution settings. A low uniform static magnetic field of approximately 34 Gauss was applied perpendicularly to the HTS microstrip line using a solenoid wound around the cryostat. The field was left on long enough for transience in the IMD to complete. Magnet current was sampled by the Python script using a digital ammeter. To begin the relaxation measurement, the static magnetic field was removed by opening a switch in the coil circuit. The drop in coil current was detected by the ammeter, prompting the Python script to begin recording timed IMD power from the spectrum analyzer. A higher sampling rate (higher resolution bandwidth) of about 8 samples per second was used in the first 1.5 seconds of relaxation where the IMD changed more rapidly, after which the sampling was slowed to about 2 per second for improved noise which was necessary as the IMD relaxed to a lower level.

These measurements were conducted for second and third order IMD at temperatures ranging from 78K to 89K. The device under test was thermal-cycled above $T_c$ prior to each run to ensure against residual magnetic flux.

In order to observe fourth and fifth order IMD above the noise floor, higher power settings of +28 dBm for $f_1$ and $f_2$ and a lower measurement temperature were needed. The $f_3$ power output was kept at +15dBm. Measurements were performed at a sample temperature of $(79.50 \pm 0.01)$K on the control sample for second, third, fourth, and fifth order IMD relaxation, and on the test sample for second, third, and fourth order IMD. Fifth order IMD was not within the sensitivity of the instruments for the test sample.

## 3. Results and Discussion

The previous study [9] of magnetic relaxation of microwave nonlinearity in YBCO resonators, which focused on temperatures closer to $T_c$, also found a slow logarithmic process and a fast exponential decay process. The logarithmic process was attributed to thermal activation over the surface barrier [6], and the exponential process was attributed to the expulsion of remanent magnetization in the superconductor after removal of the external magnetic field [7]. Measurements of the control sample in this paper replicate the previous findings at temperatures near $T_c$ and add new findings at lower temperatures. The test sample underwent relaxation that requires a different description from relaxation in the control sample.

### 3.1. Second Order IMD

Qualitatively distinct relaxation of second order IMD in various temperature ranges is shown in Figure 5 for (a) the control sample and (b) the test sample.

The general case of second order IMD in the control sample shows a fast exponential decay followed by a slow logarithmic process. This was modelled in [9] as

$$y = y_0 + A_1 e^{-t/t_1} + A_2 \log(t) . \tag{1}$$

The fit parameter $t_1$ is the exponential decay time constant. $A_1$, $A_2$, and $y_0$ set the offset and the scale of the slow and fast processes. The logarithmic description becomes unphysical at small time values due to a singularity at $t$=0. This has previously been addressed by including a time offset, but only when considering flux creep [17], which does not drive relaxation in the control sample. Here, the singularity occurred prior to beginning data collection, which made a time offset unnecessary.

In the control sample, the fast exponential process was observed at temperatures below approximately $0.894T_c$ and above approximately

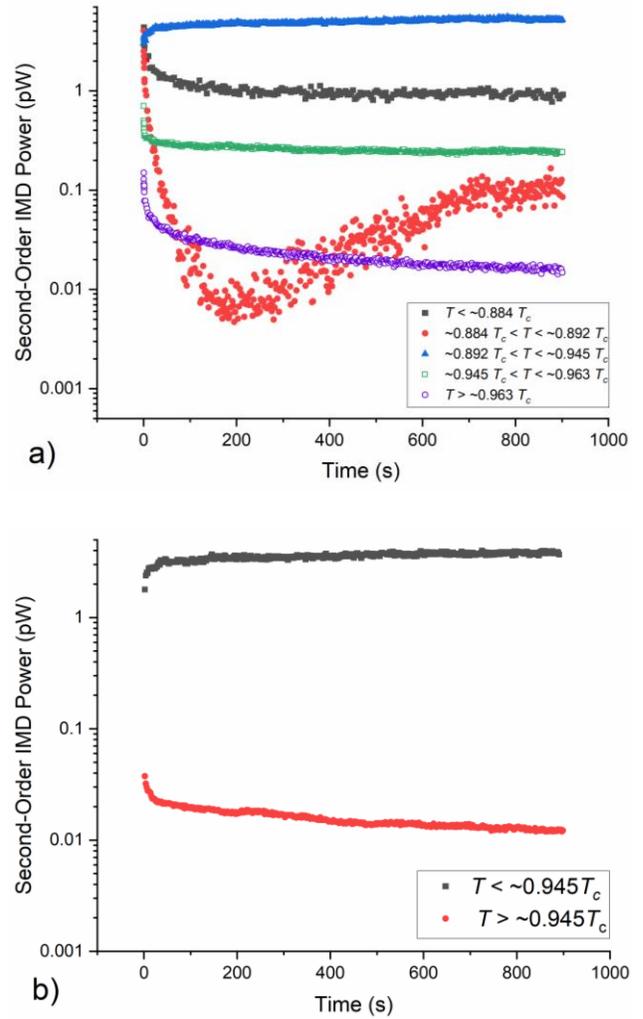

**Figure 5.** Sample relaxation measurements of second order IMD, typical of the indicated reduced temperature ranges, in a) the control sample ($T_c$=89.8 K), and b) the test sample ($T_c$=88.2 K).

$0.963T_c$. Without the time delay that is associated with depinning, remanent demagnetization proceeds quickly and exponentially upon removal of the static field, causing a drop in the level of second order IMD. Thermal activation dominates the relaxation between $0.894T_c$ and $0.963T_c$, so that the fast exponential process is masked in this range of temperatures. In this temperature range, a purely logarithmic fit has randomly scattered residuals all the way down to the first data point. If there is any fast exponential process, it cannot be detected concurrently with such a large logarithmic process characteristic of this temperature range. The decay time constant was scattered around a value of 25 seconds with no evident temperature dependence, indicating that the rate of expulsion of excess fluxons is not related to the superconducting order parameter.

The test sample did not show any exponential process within the precision of the measurement. This suggests that enough of the flux was strongly pinned to cause the transience in IM2 due to remanent demagnetization to be too weak to observe. The logarithmic thermal activation over the surface barrier was the dominant feature of the relaxation.

The process underwent changes in concavity at two temperatures, approximately $0.884T_c$ and approximately $0.945T_c$ (although the lower temperature was barely reached with the test sample). Changes of concavity manifest as sign changes of the logarithm coefficient $A_2$ in Figure 6. At the exact temperature where the concavity changes from positive to negative, there is no logarithmic process, hence no detection of Bean-Livingston type flux expulsion. The error bars from the fit for each point were smaller than the size of the data markers, so they are not shown.

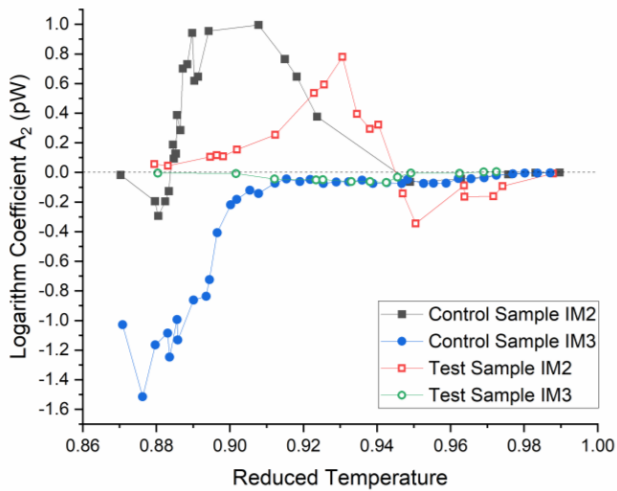

**Figure 6.** Logarithm coefficient by reduced temperature in each sample

As the sample temperature approaches $T_c$ from lower temperatures, the logarithmic process becomes indistinguishable from the scatter of the data as the strength of the IMD signal decreases. The logarithmic behavior also weakens at temperatures approximately 10K below $T_c$, where the pinning (Labusch) constant is known to increase by orders of magnitude and the pinning force becomes large [18]. It would be interesting to do these IMD decay experiments with certain iron-based superconductors where the Labusch parameter has been found by microwave measurement to be at least an order of magnitude smaller than it is for YBCO [19]. Finally, it may be possible that at even lower temperatures the short coherence length could lead again to a reduction in the pinning potential [20] and a resurgence in the logarithmic time decay of IMD.

The test sample showed only a logarithmic relaxation process. The coefficient of the logarithm is qualitatively distinct from the control sample, indicating a different physical mechanism between the fast and slow processes. There is a single inflection temperature at approximately $0.945T_c$, matching that of the control sample. As with the control sample, the logarithmic relaxation becomes weak as the temperature approaches $T_c$ from lower temperatures. Unlike the control sample, this is not seen at the low end of the available temperature range. Because of the strong pinning in the columnar defects, relaxation in the test sample is likely due to thermal activation described by the Anderson-Kim model of fluxon hopping between pinning sites [5], which results in a logarithmic fit that is distinct from the relaxation over the surface barrier which is seen in the control sample without columnar defects.

### 3.2. Third Order IMD

At nearly every temperature, third order IMD is not measurably influenced by remanent demagnetization, which is likely masked by the slow logarithmic decay. It could possibly be inside the resolution of the test, which was between 125 ms and 333 ms, depending on noise requirements. The logarithm coefficients are shown in Figure 6. Each sample showed only negative logarithm coefficients, with concave up relaxation at all tested temperatures. As the temperature of each sample approaches $T_c$ from lower temperatures, the relaxation becomes weaker as the signal strength goes to zero. The control sample relaxation coefficient for third order IMD (Figure 6) has significantly different temperature dependence when compared to the coefficient of the second order IMD. The surface barrier activation process appears to affect the second- and third order nonlinearity differently.

Below approximately $0.905T_c$, all thermally activated time development of third order nonlinearity ceases in the test sample. In this same temperature region, logarithmic relaxation is at its strongest in the control sample. The vanishing of the relaxation in the test sample may be explained by a cutoff in temperature below which thermal activation is too weak to overcome the defect pinning, meaning that the sample does not need a surface barrier potential in order to retain fluxons.

as second and third order IMD, respectively, though with weaker signal strength. Thermal activation over the surface barrier was likely the driver of the control sample relaxation at these higher orders, though the low signal power caused the logarithmic curve to be very faint relative to the scatter of the data.

At the 79.5K measurement temperature, the test sample showed no relaxation of third order IMD above the noise limit of the instrument. The even order signals showed similar behavior to the control sample, with fourth order IMD having the same relaxation curve as second order IMD, though with a weaker signal power.

The control sample and the test sample show different concavities in their even order IMD relaxation at the same temperature, likely due to the different thermally activated logarithmic relaxation processes that dominate the IMD power in the two samples. This is more evident in odd order IMD, where the control sample shows a clear logarithmic process in third order IMD while the test sample has no visible curvature due to low-temperature suppression of flux creep.

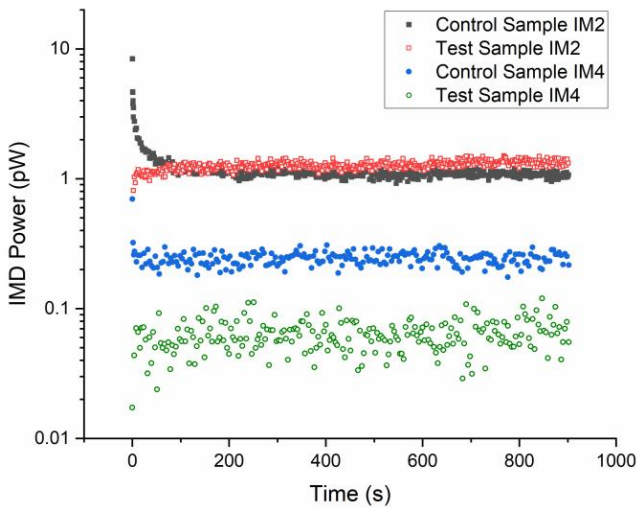

**Figure 7.** IM2 and IM4 relaxation data at 79.5 K.

### 3.3. Higher Order IMD

IMD can be generalized beyond the second and third order to higher order even and odd order IMD [21]. Higher order IMD relaxation showed similar time development as the lower orders of the same parity, as shown in Figure 7 and Figure 8. At the same temperature, fourth and fifth order IMD in the control sample relaxed with the same shape

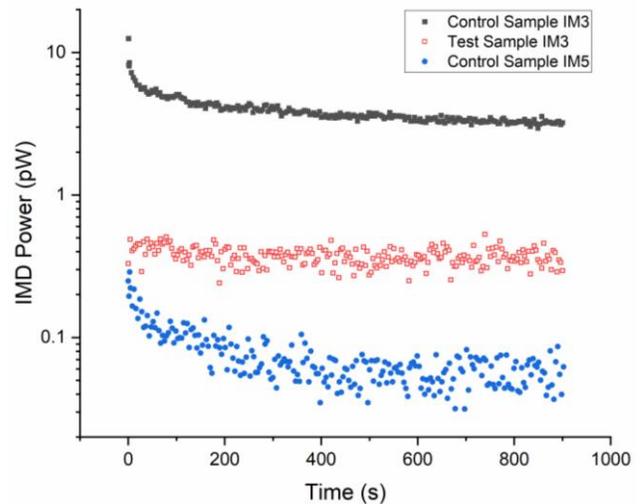

**Figure 8.** Control sample IM3 relaxation with control sample IM5 and test sample IM3 relaxation, all at 79.5 K.

## 4. Conclusion

Microwave nonlinearity in superconductors relaxes differently when fluxons are pinned by columnar defects. Remanent demagnetization is masked by the slower thermal process at all temperatures when pinning occurs, but is evident in the second order IMD, though rarely in third order IMD, of pristine material. Logarithmic time development is present in superconductors with or without columnar defects, but the different relaxation characteristics indicate that thermal depinning from the columnar defects controls the relaxation, unlike the thermal activation over the surface barrier observed in a control sample. Second order IMD and third order IMD show different relaxation behavior, while higher orders of IMD undergo relaxation with the same shape as lower orders of the same parity. Both the fast remanent demagnetization and the slow thermal activation processes influence the nonlinear microwave emissions, indicating that the flux lattice responds nonlinearly to microwave excitation.

## Acknowledgement

This material is based upon work supported by the National Science Foundation under Grant No. 1505617.

## References

[1] S.K. Remillard, H.R. Yi, A. Abdelmonem, IEEE T. Appl. Supercond 13 (2003) 3797-3802. https://doi.org/10.1109/TASC.2003.816205.
[2] I. Ciccarello, C. Fazio, M. Guccione, M. Li Vigni, Physica C 159 (1989) 769-776. https://doi.org/10.1016/0921-4534(89)90147-0.
[3] S. Lee et al., Phys. Rev. B 71 (2005) 014507. https://doi.org/10.1103/PhysRevB.71.014507.
[4] J.C. Booth et al., J. Supercond. Nov. Magn 19 (2006) 531-540. https://doi.org/10.1007/s10948-006-0126-2
[5] P.W. Anderson, Y.B. Kim, Rev. Mod. Phys 36 (1964) 39-43. https://doi.org/10.1103/RevModPhys.36.39
[6] C.P. Bean, J.D. Livingston, Phys. Rev. Lett 12 (1964) 14-16. https://doi.org/10.1103/PhysRevLett.12.14.
[7] Y. Yeshurun, A.P. Malozemoff, A. Shaulov, Rev. Mod. Phys 68 (1996) 911-949. https://doi.org/10.1103/RevModPhys.68.911.
[8] A. Agliolo Gallitto, M. Li Vigni, G. Vaglica, Physica C 404 (2004) 6-10. https://doi.org/10.1016/j.physc.2003.10.039.
[9] R.A. Huizen, S.L. Hamilton, G.T. Lenters, S.K. Remillard, IEEE T. Appl. Supercond 27 (2017) 7500205. https://doi.org/10.1109/TASC.2016.2638404.
[10] S.K. Remillard et al., Supercond. Sci. Technol 27 (2014) 095006. https://doi.org/10.1088/0953-2048/27/9/095006.
[11] G. Ghigo, F. Laviano, R. Gerbaldo and L. Gozzelino Supercond. Sci. Technol. 25 (2012) 115007, https://doi.org/10.1088/0953-2048/25/11/115007.
[12] G. Ghigo *et al.*, Scientific Reports 7 (2017) 13029, https://doi.org/10.1038/s41598-017-13303-5.
[13] D. Torsello *et al.*, Supercond. Sci. Technol. 33 (2020) 094012, https://doi.org/10.1088/1361-6668/aba350.
[14] G. Sylva, E Bellingeri, C Ferdeghini, A Martinelli, I Pallecchi, L Pellegrino, M Putti, G Ghigo, L Gozzelino, D Torsello. G Grimaldi, A Leo, A Nigro, and V Braccini, Supercond. Sci. Technol. 31 (2018) 54001.
[15] G. Ghigo et al., Phys. Rev. Lett. 121 (2018) 107001, https://doi.org/10.1103/PhysRevLett.121.107001.
[16] From Mentor Graphics, Wilsonville, OR, USA, formerly Zeeland Software.
[17] E.H. Brandt, Rep. Prog. Phys 58 (1995) 1465-1594. https://doi.org/10.1088/0034-4885/58/11/003.
[18] M. Golosovsky, M. Tsindlekht, D. Davidov, Supercond. Sci. Technol 9 (1996) 1-15. https://doi.org/10.1088/0953-2048/9/1/001.
[19] N. Pompeo, K. Torokhtii, A. Alimenti, G. Sylva, V. Braccini, and E. Silva, Supercond. Sci. Technol 33 (2020) 114006. https://doi.org/10.1088/1361-6668/abb35c
[20] T.L. Hylton, M.R. Beasley, Phys. Rev. B 41 (1990) 11669-11672. https://doi.org/10.1103/PhysRevB.41.11669.
[21] A.M. Eben et al., IEEE T. Appl. Supercond 21 (2010) 595-598. https://doi.org/10.1109/TASC.2010.2085414.